\documentstyle[galley, rotate]{mn}
\onecolumn
\input{epsf}

\def\kmin1{{\int_0^{\chi_s} \omega(\chi) d \chi }}

\def\var1{ \Big [  \int  {d^2 {\bf l} \over (2
\pi)^2 )} {\rm P} \left ( { \bf l \over r(\chi) } \right )
W^2(l\theta_0) \Big ] }
\def\var {\kappa_{\theta_0}}

\def\kmin{{\kappa_m }}
\def\e{\epsilon}
\def\l{\langle}
\def\r{\rangle}
\def\f{\left (\frac {\sigma_\epsilon^2}{2} \right )}
\def\next{\nonumber \\}
\def\mg{M_g}
\def\ma{M_{\rm ap}}
\def\ms{M_s}
\def\g{\gamma}

\title[Cosmic Shear]
{Error Estimates for Measurements of Cosmic Shear}

\author[D. Munshi and  P. Coles]{Dipak Munshi$^{1}$\thanks{e-mail: mdipak@mpa-garching.mpg.de}
and Peter Coles$^2$\thanks{e-mail: Peter.Coles@nottingham.ac.uk} \\
 $^1$Max-Planck-Institut fur Astrophysik,
Karl-Schwarzschild-Str.1, D-85740, Garching, Germany\\
$^2$School of Physics \& Astronomy, University
of Nottingham, University Park, Nottingham, NG7 2RD, United
Kingdom\\
}

\pagerange{000,000}
\begin{document}
\maketitle

\begin{abstract}
In the very near future, weak lensing surveys will map the
projected density of the universe in an unbiased way over large
regions of the sky. In order to interpret the results of studies
it is helpful to develop an understanding of the errors associated
with quantities extracted from the observations. In a
generalization of one of our earlier works, we present estimators
of the cumulants and cumulant correlators of the weak lensing
convergence field, and compute the variance associated with these
estimators. By restricting ourselves to so-called compensated
filters we are able to derive quite simple expressions for the
errors on these estimates. We also separate contributions from
cosmic variance, shot noise and intrinsic ellipticity of the
source galaxies.
\end{abstract}

\begin{keywords}
Cosmology: theory -- large-scale structure of the Universe --
Methods: analytical -- Methods: statistical
\end{keywords}

\section{Introduction}

The recent detections of weak gravitational lensing of background
galaxy images by large-scale structure (Bacon et al. 2000; van
Waerbeke et al. 2000) provide an added impetus to the development
of statistical methods for handling high-quality data from weak
lensing surveys which should provide us with valuable information
about the mass distribution in the universe (Mellier 1999;
Bernardeau 1999; Bartelmann \&  Schneider 2001). The particular
benefit of weak lensing surveys is that they permit us to probe
the distribution of underlying mass in a fashion that does not
depend on an understanding of the relationship between galaxies
and the distribution of dark matter.

Following the directions set in earlier work by Gunn (1967),
Blandford et al. (1991), Miralda-Escud\'{e} (1991) and Kaiser
(1992), most current progress in weak lensing can broadly be
divided into two  categories. Villumsen (1996), Stebbins (1996),
Bernardeau et al. (1997) and Kaiser (1987) have focussed on the
linear and quasi-linear regime by assuming a large smoothing
angle. Several other authors have developed computational
techniques to simulate weak lensing catalogs: numerical
simulations of weak lensing typically employ N-body simulations,
 through which ray tracing experiments are conducted
(Schneider \& Weiss 1988; Jarosszn'ski et al. 1990; Lee \&
Paczyn'ski 1990; Jarosszn'ski 1991; Babul \& Lee 1991;  Bartelmann
\& Schneider 1991: Blandford et al. 1991). Building on the earlier
work of Wambsganns et al. (1995, 1997, 1998) the most detailed
numerical studies of lensing have been made  by  Wambsganns, Cen
\& Ostriker (1998). Other recent studies using ray-tracing
experiments have been conducted by Premadi, Martel \& Matzner
(1998), van Waerbeke, Bernardeau \& Mellier (1998), Bartelmann et
al. (1998), Couchman, Barber \& Thomas (1998) and Jain, Seljak \&
White (2000). While the former, perturbative analyses can provide
valuable information at large smoothing angle, this approach  can
not be used to study lensing on small angular scales, because the
perturbation series involved start to diverge.

More recent studies (Hui 1999; Munshi \& Jain 2000, 2001; Munshi
\& Coles 1999, 2000; Valageas 1999a,b) have demonstrated that, in
the highly non-linear regime, it is possible to combine the
well-motivated hierarchical ansatz (Davis \& Peebles 1977; Peebles
1980; Fry 1984; Fry \& Peebles 1978; Szapudi \& Szalay 1993, 1997;
Colombi et al. 1997; Scoccimarro \& Frieman 1998; Scoccimarro et
al. 1998; Balian \& Schaeffer 1989; Bernardeau \& Schaeffer 1992;
Bernardeau \& Schaeffer 1999) with the  scaling relation for
evolution of two-point correlation functions (Hamilton et al.
1991; Nityananda \& Padmanabhan 1994; Jain, Mo \& White 1995;
Padmanabhan et al. 1996; Peacock \& Dodds 1996) to make very
accurate predictions of the statistics of the lensing convergence
field for very small smoothing angular scales. This approach
offers an analytic route to the study of small scale lensing to
complement the numerical approaches mentioned above.

The hierarchical ansatz has been used to show  that lower-order
moments such as cumulants and cumulant correlators can be modelled
very accurately. It has also been found that the probability
distribution function (PDF) and the bias associated with hot spots
in convergence maps can also be predicted very accurately using
this formalism. Higher order moments are more sensitive to the
tail of the distribution function they represent, and are
consequently more sensitive to measurement errors arising from
finite size of the sample. Although there have been many detailed
studies to quantify measurement errors for  moments of the density
field (Colombi et al. 1995; Colombi et al. 1996; Szapudi \&
Colombi 1996; Hui \& Gaztanaga 1998) similar studies for weak
lensing surveys are still lacking. Schneider et al. (1998)
proposed different estimators for extracting the variance from
convergence maps and the errors associated with them. In this
paper  we extend such results to incorporate all higher order
cumulants and their two-point counterparts the cumulant
correlators (Szapudi \& Szalay 1997) which were used in the
context of lensing by Munshi \& Coles (2000). We study the
contribution to the error involved in using these estimators by
computing their variance. We list contributions from different
sources (including the discrete nature of the source distribution,
the intrinsic ellipticities associated with source galaxies, and
the finite size of the catalogue).  In previous studies of error
estimations the higher order cumulants were assumed to be zero, as
there has not been until recently an analytic prediction for the
hierarchical parameters $S_N$  in the highly non-linear regime.
Combining our result with recent analytical prediction for $S_N$
parameters for small smoothing angles will provide an accurate way
to compute estimation errors and hence actual possibility of
measuring these quantities from observational data.

The layout of the paper is follows. In Section 2 we introduce the
estimators, and in Section 3 we explain the different types of
averaging involved in computing the dispersion and mean of these
estimators. In Section 4 we develop a diagrammatic formalism to
compute the mean and the dispersion and derive very general
expression for scatter in estimates of the cumulant correlators of
arbitrary order for an arbitrary number of points. In Section 5 we
discuss the importance of our results in a general cosmological
context. We have presented the detailed expressions for specific
lower-order moments in an appendix for easy reference.

\section{Estimators for Cumulants and Cumulant Correlators}
%
%We will be using the formalism developed by Schneider et al. (1998)
%for computation of mean and dispersion of estimators of various order.
%
%
%\begin{eqnarray}
%M_g^2 = &&\pi \theta^2 \int d^2 \theta~~ Q^2(\theta) \\
%M_s^2 = &&\pi \theta^2 \int d^2 \theta~~ Q^2(\theta) \gamma_t^2(\theta) \\
%M_{\rm ap} = && \pi \theta^2 \int d^2 \theta~~Q(\theta) \gamma_t(\theta)
%\end{eqnarray}
%
The statistics most frequently used to quantify the nature of
clustering from galaxy catalogs are the moments of various orders.
These are useful both to quantify the nature of non-Gaussianity
and also to constrain the nature of initial conditions. A
particularly useful way of combining moments is in the form of
cumulants, which have been used to quantify both galaxy clustering
and lensing surveys. Unlike the cumulants derived from galaxy
catalogs, cumulants of lensing fluctuations can also differentiate
between different cosmological models. These will be the most
useful statistical descriptors for future weak lensing surveys.

Schneider et al. (1998) have proposed the use of aperture mass
statistics based on the use of a compensated filter function $U$
to smooth the weak lensing convergence field $\kappa$ defined over
a circular patch of sky with a radius $\theta_0$:
\begin{equation}
M_{\rm ap}(\theta_0) = \int d^2 {\cal \theta} U({\cal \theta})
\kappa({\cal \theta}).
\end{equation}
This particular filter function has many useful properties which
allows us write $M_{\rm ap}$ in terms of the measured tangential
component of the shear $\gamma_t$ inside a circle of radius
$\theta$ on the sky:
\begin{equation}
M_{\rm ap}(\theta_0) = \int d^2 {\cal \theta} Q({\cal \theta})
\gamma_t({\cal \theta}),
\end{equation}
where $Q(\theta_0)$ and $U(\theta)$ are related by
\begin{equation}
Q(\theta_0) = {2 \over { \theta_0}^2} \int_0^{\theta_0} d {\cal
\theta'}U(\theta') - U(\theta_0)
\end{equation}
(Schneider et al. 1998). We will use the second definition of
$M_{\rm ap}$ in our analysis. The analysis can be extended to any
other specific form of window function, such as a top hat,
although it will no longer possible to directly relate the
smoothed convergence fields with the galaxy shear $\gamma_t$
measured observationally.

We begin by defining an estimator for the cumulants which is a
natural generalization of the lower-order estimators used by
Schneider et al. (1998):
\begin{eqnarray}
%&&M_1 = \frac{(\pi \theta_0^2)}{(n)_0}\sum_{i_1}^n Q_{i_1} \e_{i_1}, \\
%&&M_2 = \frac{{(\pi \theta_0^2)}^2}{(n)_1}\sum_{(i_1,i_2)}^n Q_{i_1} Q_{i_2}  \e_{i_1} \e_{i_2}, \\
%&&M_3 = \frac{{(\pi \theta^2)}^3} {(n)_2}\sum_{(i_1, i_2, i_3)}^n Q_{i_1} Q_{i_2} \e_{i_1} \e_{i_2} \e_{i_3}, \\
M_N = \frac{(\pi \theta_0^2)^N} {(n)_{N-1}} \left [ \sum_{(i_1,
\dots, i_N)}^n Q_{i_1} \dots Q_{i_N} \e_{i_1} \dots \e_{i_N}
\right ],
\end{eqnarray}
where $n$ is the number of galaxies in the patch of size $\pi
\theta_0^2$, the $\e_i$ are individual image ellipticities, and
$N$ is the order of the cumulant. The function $Q$ and its
relation to the compensated filter function $U$ are defined above.
In equation (4) we use the notation
\begin{equation}
(n)_N\equiv n(n-1)\ldots (n-N+1)=\frac{n!}{N!}.
\end{equation}
 We propose a family of new
estimators for cumulant correlators. Cumulant correlators were
introduced in the context of galaxy surveys by Szapudi \& Szalay
(1997). The new estimators are defined as:

\begin{eqnarray}
%&&M_{11} = \frac{(\pi \theta^2)^2}{({n_1})_0({n_2})_0}\sum_{i_1}^{n_1} \sum_{j_1}^{n_2} Q_{i_1} \e_{i_1} Q_{j_1}
%\e_{j_1}, \\
%&&M_{21} = \frac{(\pi \theta^2)^3}{({n_1})_1({n_2})_0}\sum_{(i_1,i_2)}^{n_1}
%\sum_{(j_1, j_2)}^{n_2}  Q_{i_1} \e_{i_1} Q_{i_2} \e_{i_2}  Q_{j_1}
%\e_{j_1}, \\
%&&M_{31} = \frac{(\pi \theta^2)^4}{({n_1})_2({n_2})_0}\sum_{(i_1,i_2)}^{n_1}
%\sum_{(j_1, j_2)}^{n_2}  Q_{i_1} \e_{i_1} Q_{i_2} \e_{i_2}Q_{i_3}
%\e_{i_3}  Q_{j_1} \e_{j_1}, \\
%&&M_{22} = \frac{(\pi \theta^2)^4}{({n_1})_1({n_2})_1}\sum_{(i_1,i_2)}^{n_1}
%\sum_{(j_1, j_2)}^{n_2}  Q_{i_1} \e_{i_1} Q_{i_2} \e_{i_2 }Q_{j_1}
%\e_{j_1}  Q_{j_2} \e_{j_2}, \\
M_{{N_1}{N_2}} = \frac{(\pi
\theta_0^2)^{(N_1+N_2)}}{({n_1})_{{N_1}-1}({n_2})_{{N_2}-1}} \left [ \sum_{(i_1,
\dots,i_{N_1})}^{n_1}
\sum_{(j_1,\dots, j_{N_2})}^{n_2}  Q_{i_1} \e_{i_1} \dots Q_{i_{N_1}} \e_{i_{N_1} }Q_{j_1}
\e_{j_1} \dots  Q_{j_{N_2}} \e_{j_{N_2}} \right ].
\end{eqnarray}

This approach can in principle be extended to $s$-point cumulant
correlators which are defined over $s$ different patches of the
sky where  measurements have been conducted.
\begin{equation}
M_{{N_1}\dots {N_s}} = \frac{(\pi
\theta_0^2)^{(N_1+N_2+\dots+N_s)}}{({n_1})_{{N_1}-1}({n_2})_{{N_2}-1}
\dots (n_s)_{(N_s)-1}} \left [\sum_{(i_1,
\dots,i_{N_1})}^{n_1}
\sum_{(j_1,\dots, j_{N_2})}^{n_2} \dots \sum_{(m_1,\dots, m_{N_s})}^{n_s}
\prod_{p = 1}^{N_1} Q_{i_p} \e_{i_{p} }
\prod_{q = 1}^{N_2} \e_{j_{q} }Q_{j_q} \dots
\prod_{m = 1}^{N_s} \e_{i_{m} }Q_{i_m} \right ].
%\e_{i_1} \dots Q_{i_{N_1}} \e_{i_{N_1} }Q_{j_1}
%\e_{j_1} \dots  Q_{j_{N_2}} \e_{j_{N_2}}.
\end{equation}
For detailed description of these quantities see Munshi et al.
(2000) context of galaxy surveys and Munshi \& Coles (2000) for
their weak lensing counterparts. It is well-known that these
quantities carry more information then their one point
counterparts the cumulants.

In order to be useful, the signal-to-noise ratio involved in
measurements of these quantities should be high. Our main aim in
this paper is to develop analytical results which take into
account contributions from various sources of error (or
``noise''). These include the distribution of intrinsic
ellipticities for the lensed galaxies,  shot-noise resulting from
the  discrete nature of galaxy distributions, and the finite size
of the catalogues. While the last two contributions also arise
during the analysis of projected galaxy catalogs, the  intrinsic
ellipticity distribution is a source of uncertainty unique to weak
lensing surveys.

Finite catalogue size has two principal effects. The first is that
a finite volume (or projected area) can not reveal information
about structures on a scale larger than the sample. In cosmic
microwave background studies this is sometimes, though not
entirely accurately, referred to as ``cosmic variance''. We can
generally neglect this first aspect of finite volume by using
compensated filters. The other main effect is more difficult and
involves the effect of sample boundaries or edges imposed by
sample geometry, particularly if it involves complicated masks. We
shall generally refer to this second source of error as ``finite
volume''.

We will show that the different contributions to measurement
errors of cumulant correlators are in general factorizable and can
be separated into ``pure'' terms and hybrid errors associated with
measurements of one-point cumulants. Our results are quite general
and are valid for arbitrary order and for arbitrary number of
smoothed patches. Using the rules we have developed in this paper
it will also be possible to compute the higher order moments of
errors associated with different estimators. For related
discussions, see Szapudi \& Colombi (1996) and Szapudi, Colombi \&
Bernardeau (1999).

\section{Different types of Averaging}

The total shear $\epsilon_t$ can decomposed into the part which is
due to the intrinsic source ellipticity $\epsilon_t^{(s)}$ and the
ellipticity introduced by distortion of images due to weak
lensing, which we will denote by $\gamma_t$. To proceed further we
need to consider the effects of three different types of averaging
process. First, there is the {\bf average over positions} of
source galaxies within the patches which are used to compute the
cumulant correlators. Let us denote this operator, operating on an
arbitrary statistic $M$, by ${\rm P}(M)$ where
\begin{equation}
P(M) = \left ( \prod_{i=1}^n  \int \frac {d^2 {\cal \theta}_i}{\pi
\theta_0^2} \right ) M.
\end{equation}
Second, we have {\bf the average over the distribution of
intrinsic ellipticities}. This particular source of noise is
generally assumed to be Gaussian. It is further assumed that the
ellipticities of neighbouring galaxies (in projection) do not
correlate with each other.   For a given estimator say $M$ we will
denote this average  by ${\rm G}(M)$. It will only operate on the
intrinsic ellipticity variables i.e. $\epsilon^s_{ti}$. So we can
write:
\begin{equation}
G( \epsilon_{ti} \epsilon_{tj} ) = \delta_{ij}
\frac{\sigma_{\epsilon}^2}{2}, ~~~~~~~ G( \epsilon_{ti}
\epsilon_{tj} \epsilon_{tk}) = 0, ~~~~~~~~
G(\epsilon_{ti}\epsilon_{tj}\epsilon_{tk}\epsilon_{tl}) =
\delta_{ij}\delta_{kl} \frac {\sigma^2_{\epsilon}} {2} + {\rm
cycl. perm.};~~~~~~~ G(\gamma_i\gamma_j\dots) = \gamma_i\gamma_j
\dots
\end{equation}
Finally, there is the {\bf ensemble average} over different
realizations of the sky. The ensemble averaging is commonly
denoted by $\langle \dots \rangle$. Putting these averages
together we can write the expectation value of a given statistic
$M$ for a  particular patch on the sky as
\begin{equation}
{\rm E}(M) = \langle {\rm G(P } (M)) \rangle.
\end{equation}
For a more detailed description of these operators and their
commuting properties see Schneider et al. (1998).

\section{Mean and Dispersion of Estimators}
The basic formalism we will adopt in our analysis of the bias and
dispersion of these estimators is similar to that developed by
Schneider et al. (1998). Throughout the following analysis we will
focus exclusively upon {\em compensated filters}. It was shown by
previous studies that such filters are local in nature, meaning
that  correlations between neighboring cells will be quite
negligible. As we shall see, our analytical results for the
appropriate error terms for the one-point cumulants contain only
higher-order one-point cumulants, and not two-point cumulants.
Although our results will be quite accurate for the specific case
of compensated filter functions, for other smoothing windows it is
necessary to consider the correlation between different smoothing
windows. This  introduces a  bias in our estimators in addition to
the scatter we are studying here: for a thorough discussion of the
ramifications of this, see Bernardeau et al. (1997).

We will consider several patches of the sky where measurements of
shear are performed, and we will do an averaging over galaxy
positions within each of these patches, intrinsic galaxy
ellipticity and finally an ensemble averaging over all
sky-positions. These results then will be generalized to the case
when simultaneous measurements are carried over many $s$-tuples of
patches for cumulant correlators of order $s$.

To compute the mean or dispersion of these estimators we use a
diagrammatic technique, which will simplify the computation and
will allow us to write a very general expression for the
dispersion associated with them.

\subsection{Rules}

To compute the dispersion we have to consider an identical copy of
the same patch. After expanding the multinomial expression that
results from this splitting into intrinsic and lensing-induced
shear we can express the statistic as a sum of various terms which
are just products of various combination of powers of
$\epsilon^{(s)}_t$ and $\gamma_t$ represented as a diagram as
shown in Figure 1. To compute the third order moment of errors it
is necessary to consider three copies of the same patch, and so
on. The action of the various operators as discussed above will
result in pairing of these stochastic variables which can be
computed by following the rules listed below.

\begin{figure}
\protect\centerline{
 \epsfysize = 1.5truein
 \epsfbox[5 125  477 326]
 {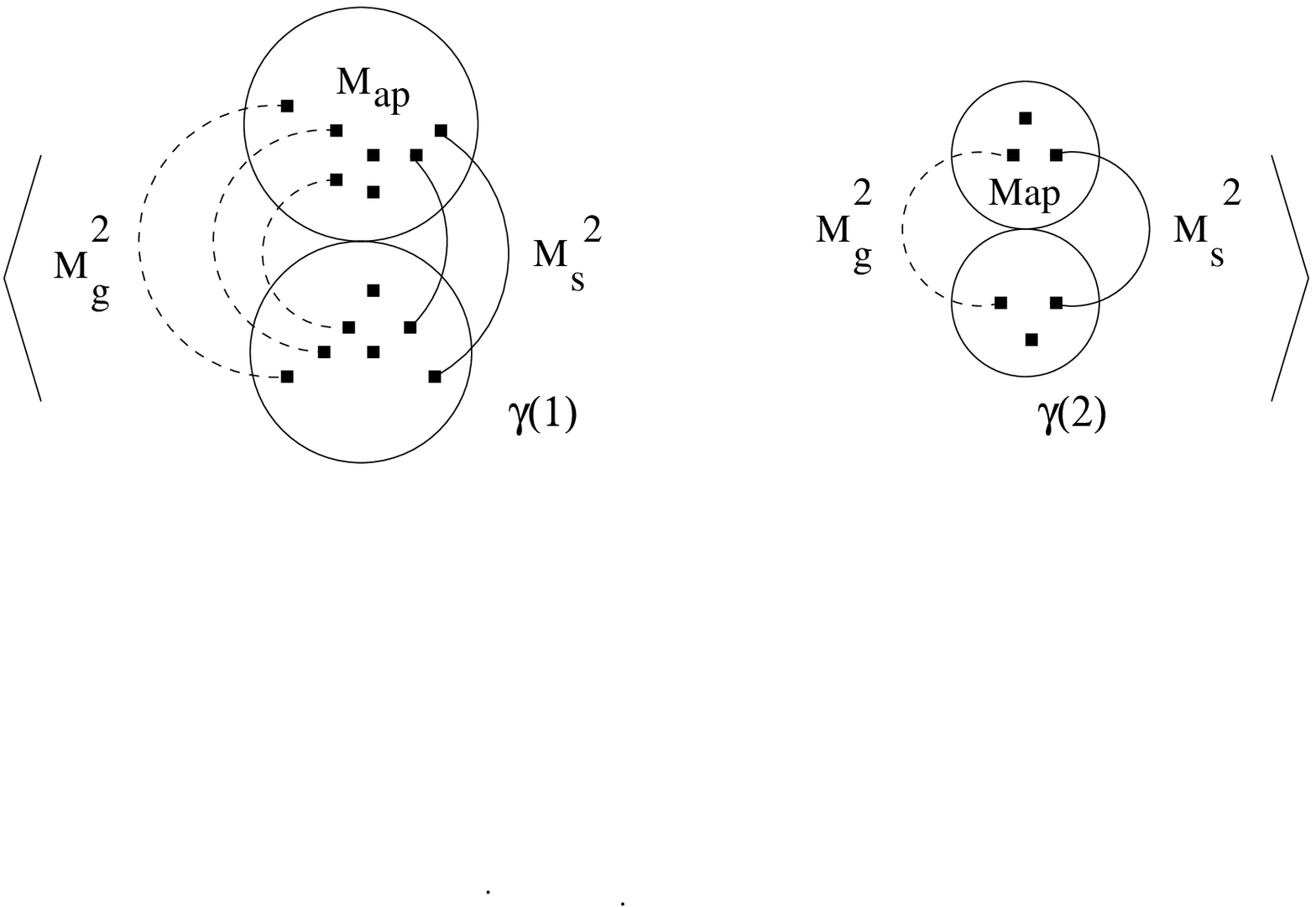} }
\caption{Computation of the variance by a diagrammatic technique.
Each circle represents one copy of the smoothed patch of the sky
in a particular direction. To compute the variance we need to
consider two copies of each patch, and so on to higher order. The
diagram shown above represents one particular term in the
expression for the dispersion. The points in each patch represents
one $\epsilon_t$, which can represent an intrinsic source
ellipticity $\epsilon^s$ or contribution due to weak lensing
$\gamma$. By assumption, source ellipticities in different patches
do not correlate but the ellipticities induced by weak lensing do
correlate. Each line joining $\gamma$ pairs has a weight factor of
$M_s$ and the lines joining $\epsilon$'s represent $M_G$. Points
which are not joined by different lines denote $M_{\rm ap}$.
Permutations of these terms in one patch of the sky do not depend
on the other patches. We have shown an error term (without
subtracting the disconnected parts) for two-point cumulant
correlators $C_{73}$ but it can be generalized trivially to
incorporate multi-point smoothed cumulant correlators.}
\end{figure}

The different pairings of $\epsilon$ will be considered between
copies of the same patch. The different types of pairing and the
rules for dealing with them are:

\begin{itemize}
\item {\bf $\gamma$ pairing}: each of these terms will contribute
one $M_s^2$ term which will denote the contribution to the error
the from discrete nature of the galaxy distribution (denoted by
solid lines in Figure 1). Here $M_s$ is defined by
\begin{equation}
M_s^2 = \pi \theta_0^2 \int_{\Omega} d^2 {\cal \theta}~~ Q^2({\cal \theta}) \gamma_t^2({\cal \theta})
\end{equation}

\item{\bf $\epsilon$ pairing}: each of these pairings will
contribute a $M_g^2$ term, which arises from the intrinsic
ellipticity of galaxies (denoted by dashed lines in  Figure 1).
\begin{equation}
 M_g^2  = \pi \theta_0^2 \int_{\Omega} d^2 {\cal \theta}~~ Q^2({\cal \theta}).
\end{equation}
The intrinsic ellipticity distribution is assumed to be Gaussian
and uncorrelated with other patches, as mentioned above.

\item {\bf No pairing of $\gamma$ or $\epsilon$} which will take
contribution from correlation term between different patches
(denoted by black dots in Figure 1). The amplitude associated with
these terms are simply $M_{\rm ap}$ which we have defined before.

\end{itemize}
%\begin{equation}
%M_{\rm ap} =  \int_{\Omega} d^2{\cal \theta}~~Q({\cal \theta})
%\gamma_t(\theta).
%\end{equation}

Total number of pairs which can be made out of $n$ objects is
$n!$: out of these pairs some of them will involve $\gamma$
pairing and some of them will correspond to $\epsilon$ pairing.

% #############################################################
\begin{figure}
\protect\centerline{
 \epsfysize = 2.5truein
 \epsfbox[35 165  277 400]
 {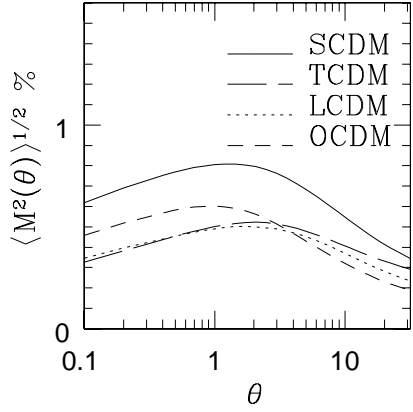} }
\caption{$M_{\rm ap}$ for various cosmological models as a
function of smoothing angle $\theta_0$ which we have used in our
error estimation. Parameters characterizing these cosmological
models can be found in Munshi \& Jain 2000.}
\end{figure}
%###########################################################

\subsection{Expectation Values}

It is not difficult to show that the estimators we proposed in
Section 2 for the cumulants and cumulant correlators are unbiased
estimators. From a diagrammatic point of view, it is clear that
there are no terms that correspond to coupling of
$\epsilon_t^{(s)}$ terms  with $\epsilon_t^{(s)}$ either from the
same patch (because no indices in the same patch can be equal) or
from two different patches (because they are not correlated). We
also note that $\gamma_t$ and $\epsilon_t^{(s)}$ are also
uncorrelated. This means that the only contribution comes from
terms in which the $\gamma_t$ are correlated. Hence we can write,
for the cumulants,
\begin{eqnarray}
E(P(A(M_N))) = \left \langle \prod_{i=1}^N\int_{\Omega} {d^2 {\cal \theta}_i} Q({\cal \theta}_i)
\gamma_t({\cal \theta}_i) \right \rangle = \l M^N_{\rm ap}\r
\end{eqnarray}
and similarly for cumulant correlators,
\begin{eqnarray}
E(P(A(M_{N_1N_2\dots}))) = \left \langle
\prod_{i=1}^{N_1}\int_{\Omega_1} {d^2 \theta_i} Q({\cal \theta}_i)
\gamma_t({\cal \theta}_i)\prod_{j=1}^{N_2}\int_{\Omega_2} {d^2 {\cal \theta}_j}
Q({\cal \theta}_j) \gamma_t({\cal \theta}_j) \dots \right \rangle =
\l M_{\rm ap}^{N_1}(\gamma_1) M_{\rm ap}^{N_2}(\gamma_2) \dots \r,
\end{eqnarray}
which shows that they are unbiased estimators of multi-point
moments.

\subsection{Dispersion}

% #############################################################
\begin{figure}
\protect\centerline{
 \epsfysize = 2.5truein
 \epsfbox[21 426  587 715]
 {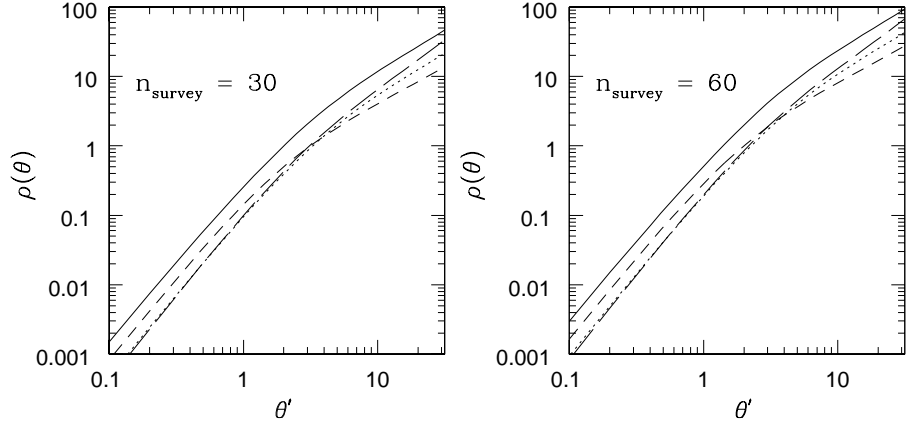} }
\caption{The parameter $\rho({\theta_0})$ which appears in our
expressions for fractional errors of various cumulants. Different
curved denote various background cosmology. Notations are same as
in previous figure. For small values of $\rho(\theta_0)$ the
errors are dominated by contributions from shot-noise where as for
large values of $\rho(\theta_0)$ it is dominated by cosmic
variance.  Fractional errors in various parameter are polynomial
in inverse $\rho$ and minima of these polynomials occur on an
arcminute scale. For left panel we have assumed 30 galaxies per
square arcminute and 60 galaxies per square arcminute for the
right panel. }
\end{figure}
%###########################################################

As we discussed above, for the computation of variances we have to
consider two copies of the same patch. Following  the rules given
above we can finally write down the expression for the dispersion
of an $N$-th order cumulant as
\begin{eqnarray}
\sigma^2(M_N) & = & \big ( (n)_{2N} \big ) \langle M_{\rm ap}^{2N}
\rangle +~~ \big ( N! \big ) \sum_{p=1}^N \big ( {}^NC_p \big )
\big ( (n)_{2N-p} \big )  \left \langle M_{\rm ap}^{2(N-p)}
M_s^{2p} \right \rangle \nonumber \\ && +~~ \big ( N! \big
)\sum_{r=1}^{N} \big ( {}^NC_r \big ) \big ( (n)_{2N-r} \big )
\left ( M_g^2 \right )^{r} \langle M_{\rm ap}^{2(N-r)} \rangle
\left ( {\sigma_\epsilon \over 2}\right )^r \nonumber \\ &&+~~\big
( N! \big ) \sum_{2 \le p+r \le N} \big ( {}^NC_r \big ) \big (
{}^{(N-r)}C_p \big ) \big ( (n)_{2n-p-r} \big ) \big ( M_g^{2}
\big )^r \langle M_{\rm ap}^{2(N-p-r)} M_s^{2p} \rangle \left (
{\sigma_\epsilon \over 2}\right )^r.
\end{eqnarray}
The symbol ${}^NC_p$ is used to denote the number of combinations
of $p$ objects taken from $N$. The first term here represents the
case when all points are $\gamma_t$ in both patches and there is
no pairing of these points within the copies of the same patch.
The second term represents only $\gamma_t$ pairing between pairs
of $\gamma_t$ from two different copies of the same patch. The
third term represents the $\epsilon_t$ pairing within the copies
of the same patch and the last terms is a mixture when some of the
couplings are $\epsilon_t$ coupling and some of them are
$\gamma_t$ coupling.

A similar analysis can be performed  for s-point smoothed cumulant
correlators of order $N_1 + \dots + N_s$. We find
\begin{eqnarray}
{\huge \sigma}^2(M_{N_1N_2\dots N_s}) & = &\Big \langle
\prod_{i=1}^s \Big \{ \big ( {(n_i)}_{2N_i} \big )  \big ( M_{\rm
ap}^{2} (\gamma_i) \big )^{N_i} + \big ( N_i! \big )
\sum_{p=1}^{N_i} \left ( {}^{N_i}C_p \right ) \left (
{(n_i)}_{2N_i-p} \right ) \left (M_{\rm ap}^2 (\gamma_i) \right
)^{(N_i-p)}\left ( M_s^2(\gamma_i) \right )^{p} \nonumber \\
&&+~~\left ( N_i! \right )\sum_{r=1}^{N_i}\left ({}^{N_i}C_r
\right ) \big ( {(n_i)}_{2N_i-r} \big ) \big ( M_g^{2} \big )^r
\left ( M_{\rm ap}^2 (\gamma_1)\right )^{(N_i-r)} \left (
{\sigma_\epsilon \over 2}\right )^r \nonumber  \\&& +~~\left (N_i!
\right ) \sum_{2 \le p+r \le N_i} \left ( {}^{N_i}C_r  \right )
\left ( {}^{(N_i-r)} C_p  \right ) \big ( (n_i)_{2n_i-p-r} \big )
\left ( M_g^{2} \right )^r  \left ( M_{\rm ap}^2 \right
)^{(N_i-p-r)} \left ( M_s^2 \right )^{p} \left ( {\sigma_\epsilon
\over 2}\right )^r \Big \} \Big \rangle.
\end{eqnarray}
The above expression is the general expression for the dispersion
which is derive using the approximations we have explained. At
each order, for a given family of cumulant correlators, the
contribution to dispersion originates due to the three different
effects explained before and their cross terms.

%############################################################

Specific expressions for the dispersion of estimators for
particular cumulants and cumulant correlators are given  in the
Appendix. Our results here are derived for the case when there is
only one field of view. In practice, however, the part of the sky
observed will be covered by several (possibly overlapping) fields
of view from which the cumulant and cumulant correlators are
estimated. For the case of compensated filter the generalization
is quite straightforward but for other filters it will require a
more rigorous analysis than is presented here. In the context of
galaxy clustering, such effects have already been explored by
Szapudi \& Colombi (1996). The effect of source clustering will
also have to be taken into account for more realistic estimation
of errors.

% #############################################################
\begin{figure}
\protect\centerline{
 \epsfysize = 2.5truein
 \epsfbox[21 414  588 717]
 {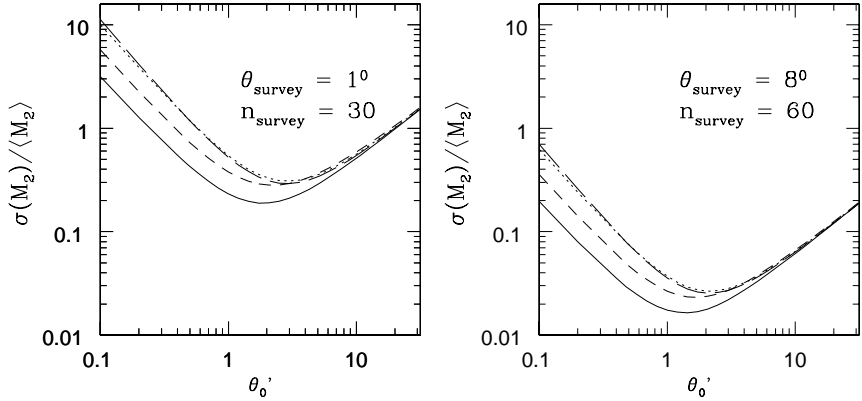} }
\caption{Estimation error for computation of $\langle M_{\rm
ap}^2\rangle$ as a function of smoothing angle $\theta_0$. A
survey area of $1^{\circ}$ is assumed for left panel and a survey
area of $8^{\circ}$ is assumed in the right panel. Different
curves denote different background cosmologies. Line style is the
same as that of Figure 2. While measurement errors are dominated
by the shot noise (which are induced due to intrinsic ellipticity
distribution of the galaxy) and the Poisson shot noise for smaller
smoothing angles. In large smoothing angles it is more dominated
by finite size of the survey area. Estimation errors for the
variance depends on the higher order moments. In particular we
found that the normalized cumulant $S_4$ appears. Unfortunately
there are no theoretical estimates for computation of $S_4$ using
compensated filters. Earlier results in the literature used
$S_4=0$, which we have used here. Clearly the fractional errors
are quite small even for moderate survey size and for large sky
coverage power spectrum estimation can be carried out with very
high precision. All our calculations are done under the hypothesis
that the correlation length scale associated with compensated
filter is quite small compared to length-scales associated with
field of view. This was demonstrated by Schneider et al. (1998).
For top-hat filters correlations between neighboring cells will
have to be taken into account. In our computation the intrinsic
eliipticity of the source galaxy $\sigma_e$ is set to be equal to
$.2$.
 }
\end{figure}
%###########################################################

% #############################################################
\begin{figure}
\protect\centerline{
 \epsfysize = 2.5truein
 \epsfbox[21 414  588 717]
 {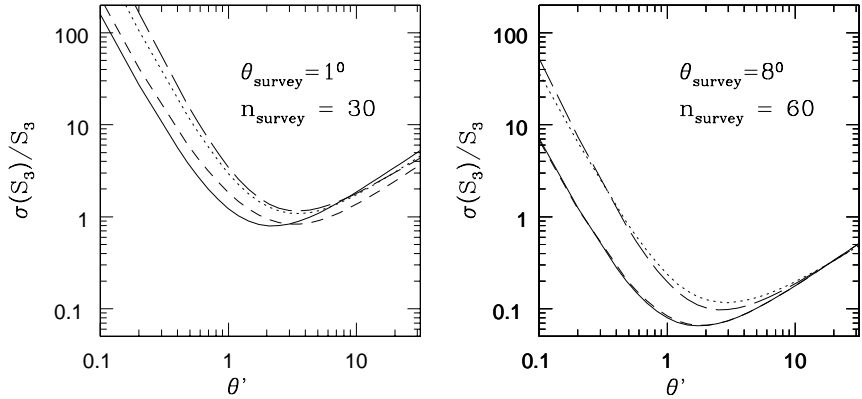} }
\caption{Estimation error for computation of $S_3$. We have assumed
that all sources are at a fixed red-shift. A survey area of radius $1^{\circ}$
(left panel) and $8^{\circ}$ (right panel) are assumed. Number density of
source galaxies $\rm n_{survey}$ in the left panel is 30 per $\rm {arminute^2}$ in the right
panel it is fixed at 60 per $\rm {arcminute^2}$. Error is plotted
as a function of the smoothing angle  $\theta_0$. Different curves
denote different background cosmolgies. Line style is same as in
previous figures. While the errors
are mostly dominated by shot noise at a very small resolution angle,
finite size of the survey are more pronounced for larger smoothing
angles. However for smoothing angular scales of few arc minutes
the fractional errors are quite small for reasonable sky coverage.
This is more interesting as non-Gaussian estimators such as $S_3$
can be used to distinguish between different cosmologies. Notice that
fractional errors do not change much while we change the background
cosmolgy. For computation of the fractional errors in $S_3$ we have
used the analytical results for compensated filters by Bernardeau
\& Valageas (2000) which are in agreement with numerical computations
by Reblinsky et al.(1999). Although these results are for smoothing
radius $\theta_0 = 4'$, the $S_N$ parameters are known not to change
much at very small angular scales where we can assume them approximately
constant for computations of error bars. }
\end{figure}
%###########################################################

\subsection{Errors associated with $S_N$ parameters}

For practical purposes the normalised cumulants and cumulant
correlators generally involve
\begin{equation}
S_N = {\langle M_{\rm ap}^N \rangle \over \langle M_{\rm ap}^2
\rangle^{N-1}}. \end{equation} These quantities are of great
importance as they can quantify the non-Gaussianity in the weak
lensing field and also can be used to study the effects of
background cosmology on weak lensing statistics. Although the
$S_N$ parameters for weak lensing surveys have already been
measured using top-hat filters, there has been not work in this
direction for computation of $S_N$ parameters using compensated
filters for high N.

Our error expression can be written in a compact form if we use
the parameters $\Sigma_N$ parameters which can be related to the
$S_N$ parameters by following expressions. These definition of
$\Sigma_N$ also include the disconnected contributions.
\begin{eqnarray}
&&\langle M_{\rm ap}^4 \rangle = S_4 \langle M_{\rm ap}^2
\rangle^3 + 3 \langle M_{\rm ap}^2 \rangle^2 =\Sigma_4 \langle
M_{\rm ap}^2 \rangle^{4/2} \nonumber \\ &&\langle M_{\rm ap}^5
\rangle = S_5 \langle M_{\rm ap}^2 \rangle^4 +10 S_3 \langle
M_{\rm ap}^2 \rangle^3 =\Sigma_5 \langle M_{\rm ap}^2
\rangle^{5/2} \nonumber \\ &&\langle M_{\rm ap}^6 \rangle = S_6
\langle M_{\rm ap}^2 \rangle^5 +15 S_4 \langle M_{\rm ap}^2
\rangle^4 + 10 S_3^2 \langle M_{\rm ap} \rangle^6 = \Sigma_6
\langle M_{\rm ap}^2 \rangle^{6/2}
\end{eqnarray}
Using the expressions derived in previous sections and in the
limit when number of source galaxies are much higher compared to
unity we can write down the following expressions for variance in
$M_3$ and $M_2$.
\begin{eqnarray}
&&\sigma^2(M_3) = \Big [ \Sigma_6 - \Sigma_3^2 + {9 \Sigma_4 \over
\rho} + {18 \over \rho^2}+{6 \over \rho^3} \Big ] \langle M_{\rm
ap}^2 \rangle^{3/2} \nonumber \\ &&\sigma^2(M_2) =   \Big
[(\Sigma_4-1) + {4 \over \rho} + {2 \over \rho^2} \Big] \langle
M_{\rm ap} \rangle^2
\end{eqnarray}
Where we have introduced the quantity $\rho$ defined by the
following expression:
\begin{equation}
\rho = {2 \langle M_{\rm ap}^2 \rangle N \over \sigma_{\epsilon^2}
G} = {1500 \pi \over G} \langle M_{\rm ap}^2 \rangle \Big (
{\theta \over  1 \rm arcmin} \Big )^2 \Big ( {.2 \over
\sigma_{\epsilon}} \Big ) \Big ( { n \over 30 \rm arcmin^{-2}}
\Big ).
\end{equation}
For small $\rho$ the errors are dominated by shot noise i.e. in
the small smoothing angle regime, while for large $\rho$ the
errors are dominated by cosmic variance. The error terms are a
polynomial function in $1/ \rho$. The minima of these polynomial
occurs on a scale of a few arcminutes, where neither cosmic
variance nor the shot noise terms are dominant for reasonable
survey strategy.

The behaviour of $M_{\rm ap}$ for various cosmologies is shown in
Figure 2. The function $\rho$ is plotted in Figure 3. Using the
results we have obtained, the scatter associated with the $S_N$
parameters can now be written in terms of these quantities:
\begin{equation}
{\sigma(S_N) \over S_N} = {\sigma(M_{\rm ap}^N) \over \langle
M_{\rm ap}^N \rangle} + (N-1){\sigma(M_{\rm ap}^2) \over \langle
M_{\rm ap}^2\rangle}
\end{equation}
For multiple fields covering the whole field of view factors of
$\sqrt N_f$ (where $N_f$ is the number of cells covering the
field) will have to be incorporated (see Schneider et al. 1998).
For example, the variance is a factor $N_f$ smaller. This, of
course, is only the case for compensated filters like those we are
studying here. Effect of over-sampling has not been studied in the
context of weak lensing and needs a dedicated analysis.

Final results for the scatter in variance and $S_3$ are plotted in
Figure 4 and Figure 5 respectively. For specific case studies we
have assumed two different hypothetical survey strategies. In one
case we assume that the radius of the survey is $1^{\circ}$ in the
second case it is fixed at $8^{\circ}$. The number density of
source galaxies in the first case is assumed to be 30
arcmin$^{-2}$ and twice that value in the second case. The
parameter $\sigma_{\epsilon}$ which characterizes the intrinsic
ellipiticity distribution of galaxies is fixed at $.2$ in both
cases. While the first case is representative of various ground
based surveys with less exposure time which are currently in
progress, the second case is more idealistic and hope to be
achieved by future weak lensing survey with long exposure time
probably from outside earths atmosphere. In both cases the errors
are minimum for arcminute scale. Our study shows that skewness
parameters $S_3$ can be measured very accurately from second type
of surveys although signals might be detectable even in first case
with considerable uncertainty resulting in large error bars. This
underlines the importance of future weak lensing surveys from
cosmological perspective.

Direct determination of higher order $S_N$ parameters for the case
of weak lensing surveys has so far only been reported by
Bernardeau et al. (2002).  Higher moments become more difficult to
measure because better statistics are needed. As the expression
for the error in such quantities contain $S_N$ parameters of
higher order we need to make certain order-of-magnitude
approximations to make progess in estimating errors. Schneider et
al. (1998) assumed that $\Sigma_4=3$. In addition to that we have
assumed that $\Sigma_6=\Sigma_3^2$ for error expression in $S_3$.
A more accurate error estimation will have to wait till we can
compute these quantities analytically. However any analytical
results will have to be checked with numerical result for which we
need to have a better understanding of their error properties. Our
results are only a first step in this cycle and will have to be
refined by future studies.

We stress again that our derivation is only exact for the  case of
compensated filters when neighboring smoothed regions are not
correlated. For the case of other filter functions such as the
top-hat filter a more general treatment is necessary. We plan to
present such results elsewhere. We have also assumed that the
source redshifts are determined with high accuracy. Lack of
knowledge of source redshifts is a potential source of additional
error which too needs to be taken into account for more realistic
estimation of cosmological parameters from future weak lensing
surveys.

\section{Discussion}

It has been the purpose of this paper to estimate the errors
involved with the extraction of statistical information from weak
lensing surveys. We have focussed on the cumulants and cumulants
correlators, which are normalized moments of the smoothed one- and
two-point probability distribution functions. These quantities are
widely used to quantify the statistical nature of clustering of
the mass distribution in the study of galaxy surveys. In this
context, estimators of these statistics are prone to error from
finite catalogue size and Poisson (discreteness) effects. The
application of similar methods to weak lensing studies is clearly
appropriate, but introduces an additional source of error. This
paper allows for these additional error terms.

Various estimators have already been proposed for the computation
of the variance and skewness of cosmic shear smoothed with a
particular window function. We have also generalized these
suggestions  to statistics of arbitrary order via a generalized
estimator which estimates the statistics of smoothed convergence
field in an unbiased way. We have also proposed a new set of
estimators which are useful for measuring cumulant correlators,
and which are natural generalizations of their one-point
counterparts. We have shown that our estimators constitute a
family of unbiased estimators for cumulant correlators.

We have also computed the dispersion of these estimators, which is
essential to determine the signal-to-noise ratio associated with
them. We have found compact expressions for the dispersions for
arbitrary order and also for arbitrary number of points. We have
been also able to separate contributions from different sources of
noises such as the finite size of the galaxy catalog, finite width
of ellipticity distribution of source galaxies and Poisson noise
due to finite number of source galaxies in the field of view. Our
results  are valid for both large and small smoothing angles. In
case of large smoothing angles, where perturbative calculations
are still valid, one need to use the quasi-linear values of $S_N$
and $C_{pq}$ parameters associated with the expressions for finite
volume corrections. For smaller smoothing angles we have to
replace these numbers by forms suitable for the highly non-linear
regime, which have already been computed by several authors
recently based on the hierarchical {\em ansatz} (Hui 1999; Munshi
\& Jain 2000, 2001; Munshi 2000, Valageas 2000a,b).

The effect of source clustering, which we have ignored, will also
introduce corrections terms in the measurement of lower order
moments. We have also ignored the effect due to lens coupling.
Some of these issues have been studied by Bernardeau et al. (1997)
and Schneider et al. (1998) with the result that such corrections
seem to be negligible, at least in the quasi-linear regime.
Studying the effects of source clustering using ray-tracing
experiments in the highly non-linear regime is difficult because
most such simulations propagate light rays backward: the source
position is consequently left arbitrary and determined only by
lensing due to the intervening mass.

The validity of the Born approximation, which underpins lensing
calculations, has also been studied in the quasi-linear regime. It
has been shown that corrections arising from higher-order terms in
the photon propagation equation are negligible in quasi-linear
regime. Similar conclusions have also been found to be valid in
the highly non-linear regime by comparing ray-tracing simulation
against analytical results using hierarchical ansatz. (Hui 1999;
Munshi \& Jain 2000, 2001; Munshi 2000; Valageas 2000a,b).

We have also assumed that the galaxy intrinsic ellipticities are
not correlated but it may be possible that the galaxies are not
randomly oriented and there may be a a coherent alignment due to
the geometry of the large-scale structure in which they are
embedded (e.g. Heavens, Refregier \& Heymans 2000). So far however
no convincing observations of nearby structures have indicated
that such an alignment exists (e.g. Mellier 1999) although several
attempts have been made to unearth one.

\section*{Acknowledgment}
DM was supported by a Humboldt Fellowship at the Max Planck
Institut fur Astrophysik when this work was performed. It is a
pleasure for DM to
thank Bhuvnesh Jain, Francis Bernardeau, Patric Valageas. Katrin
Reblinsky and Alexandre Refregier for many helpful discussions.

\appendix

\section{Explicit Expressions for Error Terms}

\subsection{Cumulants}

Error terms for the one-point smoothed cumulants involve three
contributions. For example, for the lowest order cumulant,
\begin{equation} \sigma^2(M_1) =~~~~~~ { 1\over
n^2} \left [ n(n-1) \langle M_{\rm ap}^2 \rangle + n \langle
M_s^2\rangle + nM_g^2 \f \right ] - \l M_{\rm ap}\r^2.
\end{equation}
The first of these contributions derives from finite volume
effects included in the higher order moments such as $M_{\rm ap}$.
For example
 the dispersion in variance depends on the fourth moment, and the
dispersion in skewness will similarly depend on the sixth order
moment. The terms denoted by $M_s^2$ are related to the fact that
we have a finite number of galaxies: this will vanish if we take
the limiting case of infinite number of galaxies. The terms with
$M_g^2$ are due to the finite width of the intrinsic galactic
ellipticity distribution and would vanish in the limit in which
there are an infinite number of galaxies in the survey.

For the higher-order cumulants there are differences. Unlike the
expression for the dispersion of $M_1$, we will have mixed terms
in the expression for the dispersion in $M_2$ which are denoted by
various products of $M_s$, $M_{\rm ap}$ and $M_g$. The following
expression  was derived by Schneider et al. (1998):
\begin{eqnarray}
\sigma^2(M_2) & = &{1 \over n^2(n-1)^2} \Big [ \Big \{ n(n-1)\dots
(n-3) \l M_{\rm ap}^4 \r + 4n(n-1)(n-2) \l M_s^2 M_{\rm ap}^2 \r +
2n(n-1) \l(M_s^2)^2 \r \Big \} \next && + \Big \{ 4 n(n-1)(n-2)
M_g^2 \l M_{\rm ap}^2 \r + 4n(n-1)(n-2) M_g^2 \l M_s^2 \r \Big \}
\f^2 + 2n(n-1)M_g^2 \f^2 \Big ] - \l M^2_{\rm ap} \r^2.
\end{eqnarray}

At third order we get the following expression for dispersion of $M_3$
\begin{eqnarray}
\sigma^2(M_3) & = & {1 \over n^2(n-1)^2(n-2)^2} \Big [ \Big \{
n(n-1)\dots(n-5)\langle M_{\rm ap}^6 \rangle +
9n(n-1)\dots(n-4)\langle M_{\rm ap}^4 M_s^2 \rangle \next &&+
18n(n-1)\dots(n-3) \langle M_{\rm ap}^2 (M_s^2)^2 \rangle \Big \}
\next &&+ \Big  \{ 9n(n-1)(n-2) M_g^2 \l (M_s^2)^2 \r  + 18
n(n-1)\dots(n-3)M_g \l M_s^2 M_{\rm ap}^2 \r  +  9
n(n-1)\dots(n-4) M_g^2 \l M_{\rm ap}^2 \r \Big \} \f \next && +
\Big \{ 18n(n-1)(n-2) M_g^2 \l M_s^2 \r  + 18n(n-1)\dots(n-3)
M_g^2 \l M_{\rm ap}^2 \r \Big \}\f^2 \next && + 6n(n-1)(n-2)M_g^3
\f^3 \Big \} \Big ] - \l M^3_{\rm ap}\r^2
\end{eqnarray}
Notice that we can write (Fry 1984) $\langle M_{\rm ap}^6 \rangle
= S_6 \l M^2_{\rm ap} \r_c^5 + 15 S_4 \l M^2_{\rm ap}\r_c^4 +
10S_3^2 \l M^2_{\rm ap} \r_c^4 + 15 \l M^2_{\rm ap}\r_c^3$. For
top-hat window functions we have analytic expressions for all
lower order $S_N$ parameters which can be used to estimate the
effects of  finite volume in the highly non-linear regime. For
other window functions there is no such analytical expressions.

Generally speaking, for large values of $n$, only the dominant
contributions are considered and the rest are neglected. On the
smaller angular scales the finite width of the galaxy ellipticity
distribution dominates, and for very large smoothing angles it is
the finite volume effect that dominates. Hence one can often
neglect the terms containing $M_s$ altogether.

At fourth order we get
\begin{eqnarray}
\sigma^2(M_4) & = & {1 \over n^2(n-1)^2(n-2)^2(n-3)^2} \Big [ \Big
\{ n(n-1)\dots(n-7)\langle M_{\rm ap}^8 \rangle +
16n(n-1)\dots(n-6)\langle M_{\rm ap}^6 M_s^2 \rangle \next &&+
72n(n-1)\dots(n-5) \langle M_{\rm ap}^4 (M_s^2)^2 \rangle +
96n(n-1)\dots(n-4) \langle M_{\rm ap}^2 (M_s^2)^3 \rangle +
24n(n-1)\dots(n-3) \langle (M_s^2)^4 \rangle\Big \} \next &&+ \Big
\{ n(n-1)\dots(n-6) M_g^2 \l (M_{\rm ap})^6 \r  + 3
n(n-1)\dots(n-4)M_g^2 \l (M_s^2) M_{\rm ap}^4 \r  \next && + 3
n(n-1)\dots(n-6) \l M_g^2 \l (M_s^2)^2 M_{\rm ap}^2 \r +
n(n-1)\dots(n-3)G^2 \l (M_s^2)^3 \r \Big \} \f \next && + \Big \{
144n(n-1)(n-2) \dots (n-5) (M_g^2)^2 \l M_{\rm ap}^4 \r  +
288n(n-1)\dots(n-4) (M_g^2)^2 \l M_{\rm ap}^2 M_s^2 \r + \next
&&144n(n-1)\dots(n-3) (M_g^2)^2 \l (M_s^2)^2 \r \Big \}\f^2 \next
&& + 24n(n-1)(n-2)(n-3)M_g^4 \f^4 \Big \} \Big ] - \l M^4_{\rm ap}\r^2.
\end{eqnarray}

In the statistical study of large scale distribution of galaxies
generally cumulants are normalized by dividing them with suitable
power of two-point cumulant or by the variance, for example in the
construction of the $S_N$ parameters. Since both the numerator and
denominator are both affected by errors we have discussed above
they will introduce a ratio bias as discussed by Hui \& Gaztanaga
(1999).

Finally in this section we mention that the above results do not
depend on the scale of non-linearity probed by weak lensing, but
the appropriate values of $S_N$ must be used. For example, when
large smoothing angles are considered we should use the
quasi-linear values of $S_N$ parameters of the convergence map.

\subsection{Cumulant Correlators}

To compute the cumulant correlators we consider two patches in the
sky in the direction of $\gamma_1$ and $\gamma_2$. The results are
very  similar to the case of one-point smoothed cumulants. As
before the errors associated with measurements can be
differentiated in three  types, i.e. the finite volume effects,
errors due to intrinsic ellipticity of the source galaxies and
errors associated with finite number of galaxies. The lowest order
cumulant correlator is of course the smoothed two-point
correlation function. The error associated with its measurement
can be expressed as
\begin{eqnarray}
\sigma^2(M_{11}) & = & \frac {1} {n_1^2n_2^2}\Big [ \Big \{
n_1(n_1-1)n_2(n_2-1) \l M_{\rm ap}^2(\gamma_1)M_{\rm
ap}^2(\gamma_2) \r + n_1n_2(n_2-1) \l M_s^2(\gamma_1) M^2_{\rm
ap}(\gamma_2)\r \next && +n_2n_1(n_1-1) \l M_s^2(\gamma_2)
M^2_{\rm ap}(\gamma_1)\r +n_1n_2 \l M_s^2(\gamma_1)
M^2_{s}(\gamma_2)\r  \Big \} \next &&+ \Big \{
n_1n_2(n_2-1)M_g^2\l M^2_{\rm ap}(\gamma_2)\r  +
n_2n_1(n_1-1)M_g^2\l M^2_{\rm ap}(\gamma_1)\r + n_2n_1M_g^2\l
M^2_s(\gamma_2) \r +n_2n_1M_g^2\l M^2_s (\gamma_1)  \r \Big \} \f
\next && +  n_1 n_2 M_g^2 \f^2 \Big ] - \l M_{\rm ap} (\gamma_1) M_{\rm ap}
(\gamma_2) \r^2.
\end{eqnarray}
We have taken account that the number of galaxies in two different
patches can be different. Notice that error now contains terms
which are mainly the correlation of measurement errors in two
different patches.

Terms of higher order begin with $M_{21}$, which can be given in
the form
\begin{eqnarray}
\sigma^2(M_{21})  & = & \frac{1} {n_1^2n_2^2(n_2-1)^2} \Big [ \Big
\{ 4n_1(n_1-1)(n_1-2)n_2(n_2-1) \l \ms^2(\gamma_1)
\ma^2(\gamma_1)\ma^2(\gamma_2) \r \next &&+
n_1(n_1-1)(n_1-2)(n_1-3)n_2\l \ma^4(\g_1)\ms^2(\g_2)\r +
2n_1(n_1-1)n_2(n_2-1) \l \ms^2(\gamma_1)\ma^2(\gamma_2)\r \next &&
+4n_1(n_1-1)(n_1-2)n_2 \l \ms^2(\gamma_1) \ma^2(\gamma_1)
\ms^2(\g_2) \r + n_1(n_1 -1)(n_1 -2)(n_1 -3)n_2(n_2 -1) \l
\ma^4(\g_1)\ma^2(\g_2)\r \next &&+ 2n_1(n_1-1)n_2 \l \ms(\g_1)^2
\ms(\g_2)\r \Big \} \next && + \Big \{
4n_1(n_1-1)(n_1-2)n_2(n_2-1)\mg^2(\g_1)\l \ma^2(\g_1) \ma^2(\g_2)
\r  + 4n_1(n_1-1)n_2(n_2-1) \mg^2(\g_1) \l \ms^2(\g_1) \ma^2(\g_2)
\r  \next &&+4n_1(n_1-1)n_2(n_2-1) \mg^2(\g_1) \l \ma^2(\g_1)
\ms^2(\g_2) \r  +4n_1(n_1-1)n_2 \mg^2(\g_1) \l \ms^2(\g_1)
\ms^2(\g_2) \r \Big \} \f \next && + \Big \{
n_1(n_1-1)(n_1-2)(n_1-3)n_2 \mg^2(\g_2) \l\ma^4(\g_1)\r + 4n_1(n_1
-1)(n_1 -2)n_2 \mg^2 \l\ms^2(\gamma_2) \ma^2(\g_1) \r \next &&+
2n_1(n_1-1)n_2 \mg^2(\gamma_1) (\ms^2(\g_1))^2  \Big \} \f \next
&&+ \Big \{ 4n_1(n_1-1)(n_1 -2)n_2 \mg^2(\gamma_1) \mg^2(\gamma_2)
\l \ma^2(\g_1) \r  + 4n_1(n_1-1)n_2 \mg^2(\gamma_1)
\mg^2(\gamma_2)  \l \ms^2(\g_1) \r \next && +
2n_1(n_1-1)n_2(n_2-1) (\mg^2(\g_1))^2 \l \ma^2(\g_2) \r +
2n_1(n_1-1)n_2(\mg^2(\g_2))^2 \l \ms^2(\g_2) \r \Big \} \f^2 \next
&& + 4 n_1(n_1 -1)n_2 (\mg^2(\gamma_1))^2 \l \ms^2(\gamma_2) \r
\f^3 \Big ] - \l M^2_{\rm ap} (\gamma_1) M_{\rm ap}
(\gamma_2) \r^2.
\end{eqnarray}

These expressions are cumbersome, but worth listing because these
are the cumulant correlators most likely to be measurable. In the
above expressions we used
\begin{eqnarray}
M_s^2(\g_i) & = & \pi \theta^2 \int_{\Omega_i} d^2 \theta~~
Q^2(\theta) \gamma_t^2(\theta) \next M_g^2(\g_i)  & = & \pi
\theta^2 \int_{\Omega_i} d^2 \theta~~ Q^2(\theta). \next M_{\rm
ap}(\g_i) & = & \int_{\Omega_i} d^2 \theta~~Q(\theta)
\gamma_t(\theta).
\end{eqnarray}

In deriving the above expressions for measurement errors in
two-point cumulant correlators we have again assumed that the
intrinsic ellipticities of  galaxies do not cross correlate among
different patches. At increasingly higher orders the expressions
for error contribution becomes more complicated, although as
explained above for large number of galaxies we can take the $n_1
\rightarrow \infty$ and $ n_2 \rightarrow \infty$ limit which will
simplify these expressions.

As in the case of cumulants normalized cumulant correlators can be
derived by dividing cumulant correlators by suitable powers of
variance within these patches and the correlation among these
patches (see Munshi \& Coles 2000). Normalized cumulant
correlators are also denoted by $C_{pq}$ and in addition to the
errors we have already discussed we will have ratio bias too.

The formalism we have developed above can also be extended to
compute the skewness associated with these estimators. Statistical
studies for galaxy distribution have already shown that estimated
values of these moments are more likely to have a lower value then
its mean (Szapudi \& Colombi 1996). This will mean that the
probability  distribution of these estimators are skewed. The
skewness associated with these estimators can also be computed
using the procedure outlined above and we hope to present a
detailed analysis elsewhere; see Szapudi et al. (2000) for
comments in a similar vein.

The results presented above are valid for only one particular
patch of the sky for computation of cumulants and a single pair of
patches for measurements of cumulant correlators. However it is
straightforward to generalize the above results for large number
of patches or pairs of patches which will help to increase the
signal to noise ratio (see Schneider et al. 1998).

Detailed expressions for many-point cumulant correlators can also
be obtained using the formalism developed here. The general
expression we have presented in the text are valid for an
arbitrary number of points and do not depend directly on the
filter function used. In deriving the above results we have
assumed that the two patches have same size and hence the same
variance, but they might contain different
number of galaxies, i.e. $n_1$ and $n_2$. It is not  difficult to
generalize the results to the case when the variance in these two
patches are also different. We aim to present numerical results
for specific filters which arise from above results and their
comparison against simulated noisy convergence maps in the near
future.

%##########################################################################

%############################################################################


\begin{thebibliography}{}
\bibitem{} Babul A., Lee M.H., 1991, MNRAS, 250, 407
\bibitem{}Bacon D.J., Refregier A., Ellis R.S., 2000, MNRAS, 318,
625
\bibitem{BaSa} Balian R., Schaeffer R., 1989, A\& A, 220, 1
\bibitem{} Bartelmann M., Huss H., Colberg J.M., Jenkins A., Pearce F.R., 1998, A\&A, 330, 1
\bibitem{} Bartelmann M., Schneider P., 2001, Phys. Rep., 340, 292
\bibitem{B99} Bernardeau F., 1999, in Lachieze-Rey, ed., Proc.
Carg\`{e}se Summer School on Theoretical and Observational
Cosmology. Kluwer, Dordrecht.
\bibitem{} Bernardeau F., Schaeffer R., 1992, A\& A, 255, 1
\bibitem{} Bernardeau F., Schaeffer R., 1999, A\& A, 349, 697
\bibitem{} Bernardeau F., van Waerbeke L., Mellier Y., 1997, A\& A,
322, 1
\bibitem{} Bernardeau F., van Waerbeke L., Mellier Y., 2002, A\& A,
389, 28
\bibitem{} Blandford R.D., Saust A.B., Brainerd T.G., Villumsen
J.V., 1991, MNRAS, 251, 600
\bibitem{} Boschan P., Szapudi I., Szalay A.S., 1994, ApJS, 93, 65
\bibitem{} Colombi S., Bernardeau F., Bouchet F.R., Hernquist L.,
1997, MNRAS, 287, 241
\bibitem{} Colombi S., Bouchet F.R., Schaeffer, R., 1995 ApJS, 96,
401
\bibitem{}  Colombi S., Bouchet F.R., Hernquist L., 1996, ApJ,
465, 14
\bibitem{} Couchman H.M.P., Barber A.J., Thomas P.A., 1998, MNRAS,
308, 180
\bibitem{} Davis M., Peebles P.J.E., 1977, ApJS, 34, 425
\bibitem{} Fry, J.N., 1984, ApJ, 279, 499
\bibitem{} Fry, J.N., \& Peebles, P.J.E., 1978, ApJ, 221, 19
\bibitem{} Groth, E., \& Peebles, P.J.E., 1977, ApJ, 217, 385
\bibitem{} Gunn, J.E., 1967, ApJ, 147, 61
\bibitem{} Hamilton A.J.S., Kumar, P., Lu, E. \& Matthews, A., 1991,
ApJ, 374, L1
\bibitem{} Heavens A.F., Refregier A., Heymans C., 2000, MNRAS,
319, 649
\bibitem{} Hui L., 1999, ApJ, 519, 622
\bibitem{} Hui L., Gaztanaga E., 1999, ApJ, 519, 622
\bibitem{} Jain B., Mo H.J., White S.D.M., 1995, MNRAS, 276, L25
\bibitem{} Jain B., Seljak U., 1997, ApJ, 484, 560
\bibitem{} Jain B., Van Waerbeke L., 2000, ApJ, 530, L1
\bibitem{} Jain B., Seljak U., White S.D.M., 2000, ApJ, 530, 547
\bibitem{} Jaroszyn'ski M., Park C., Paczynski B., Gott J.R.,
1990, ApJ, 365, 22
\bibitem{} Jaroszyn'ski M. 1991, MNRAS, 249, 430
\bibitem{} Kaiser N., 1992, ApJ, 388, 272
\bibitem{} Kaiser N., 1998, ApJ, 498, 26
\bibitem{} Limber D.N., 1954, ApJ, 119, 665
\bibitem{} Lee M.H., Paczyn'ski B., 1990, ApJ, 357, 32
\bibitem{} Miralda-Escud\'{e} J., 1991, ApJ, 380, 1
\bibitem{} Mo H., White S.D.M., 1996, MNRAS, 282, 347
\bibitem{} Munshi D., Bernardeau F., Melott A.L., Schaeffer R.,
1999, MNRAS, 303, 433
\bibitem{} Munshi D., Melott A.L, 1998, preprint, astro-ph/9801011
\bibitem{} Munshi D., Coles P., Melott A.L., 1999a, MNRAS, 307, 387
\bibitem{} Munshi D., Coles P., Melott A.L., 1999b, MNRAS, 310, 892
\bibitem{} Munshi D., Melott A.L., Coles P., 1999c, MNRAS, 311, 149
\bibitem{} Munshi D., Coles P., 2000, MNRAS, 313, 148
\bibitem{} Munshi D., Coles P., 2002, MNRAS, 329, 797
\bibitem{} Munshi D., Jain B., 2000, MNRAS, 318, 109
\bibitem{} Munshi D., Jain B., 2001, MNRAS, 322, 107
\bibitem{} Munshi D., Jain B., 2000, MNRAS, in preparation
\bibitem{} Mellier Y., 1999, ARA\& A, 37, 127
\bibitem{} Nityananda R., Padamanabhan T., 1994, MNRAS, 271, 976
\bibitem{} Padmanabhan T., Cen R., Ostriker J.P., Summers, F.J., 1996,
ApJ, 466, 604
\bibitem{} Peebles P.J.E., 1980, {\em The Large Scale Structure of the
Universe}. Princeton University Press, Princeton
\bibitem{} Peackock J.A., Dodds S.J., 1996, MNRAS, 280, L19
\bibitem{} Premadi P., Martel H., Matzner R., 1998, ApJ, 493, 10
\bibitem{} Press W.H., Schechter P. 1974, ApJ, 187, 425
\bibitem{} Reblinsky K.,  Kruse G., Jain B., Schneider P., 1999,
A\& A, 351, 815
\bibitem{} Schneider P., Weiss A., 1988, ApJ, 330,1
\bibitem{} Schneider P., van Waerbeke L., Jain B., Kruse G., 1998,
MNRAS, 296, 873, 873
\bibitem{} Scoccimarro R., Colombi S., Fry J.N., Frieman J.A., Hivon E.,
Melott A.L., 1998, ApJ, 496, 586
\bibitem{} Scoccimarro R., Frieman J., 1998, ApJ, 520, 35
\bibitem{} Stebbins A., 1996, preprint, astro-ph/9609149
\bibitem{} Szapudi I., Colombi S., 1996, ApJ, 470, 131
\bibitem{} Szapudi I., Colombi S., Bernardeau F., 1999, MNRAS, 310,
428
\bibitem{} Szapudi I., Colombi S., Jenkins A., Colberg J., 2000,
MNRAS, 313, 725
\bibitem{} Szapudi I., Szalay A.S., 1993, ApJ, 408, 43
\bibitem{} Szapudi I., Szalay A.S., 1997, ApJ, 481, L1
\bibitem{} Szapudi I., Szalay A.S., Boschan P., 1992, ApJ, 390, 350
\bibitem{} Valageas P., 2000a, A\& A, 354, 767
\bibitem{} Valageas P., 2000b, A\& A, 356, 771
\bibitem{} van Waerbeke L., Bernardeau F., Mellier Y., 1999,
A\& A, 342, 15
\bibitem{} van Waerbeke, L., et al. 2000, A\& A, 358, 30
\bibitem{} Villumsen J.V., 1996, MNRAS, 281, 369
\bibitem{} Wambsganss J., Cen R., Ostriker J.P., 1998, ApJ, 494,
298
\bibitem{} Wambsganss J., Cen R., Xu G., Ostriker J.P., 1997, ApJ, 494,
29
\bibitem{} Wambsganss J., Cen R., Ostriker J.P., Turner E.L.,
1995, Science, 268, 274
\bibitem{} Wang Y., 1999, ApJ, 525, 651
\end{thebibliography}
\end{document}